# Analysis of the Petersen Diagram of Double-Mode High-Amplitude δ Scuti Stars

**Riccardo Furgoni**
*Keyhole Observatory MPC K48, Via Fossamana 86, S. Giorgio di Mantova (MN), Italy, and AAMN Gorgo Astronomical Observatory MPC 434, S. Benedetto Po (MN), Italy; riccardo.furgoni@gmail.com*

**Abstract**   I created the Petersen diagram relative to all the Double Mode High Amplitude δ Scuti stars listed in the AAVSO's International Variable Star Index up to date December 29, 2015. For the first time I noticed that the ratio between the two periods P1/P0 seems in evident linear relation with the duration of the period P0, a finding never explicitly described in literature regarding this topic.

### 1. Introduction

Within the wide range of pulsating variables there is a group with a very large population: the δ Scuti stars. It is a very heterogeneous group composed of stars with radial and non-radial pulsations and normally small amplitudes variations. Nevertheless a small group of stars represents a special sub-type: the Double-Mode High-Amplitude δ Scuti stars (from now HADS(B)), characterized by pulsations of higher intensity and a ratio between the period of the fundamental mode (P0) and first overtone (P1) around the value 0.77. The ratio between the two periods has been for a long time the main parameter to identify a HADS(B) compared to a simple HADS. But is there a more complex relation between P0 and P1/P0? In recent years the two major contributions that have attempted to find an extensive model capable to explain the link between the period of the fundamental mode (usually between 0.05 and 0.25 day) and the period ratio P1/P0 were those by Petersen and Christensen-Dalsgaard (1996) and by Poretti *et al.* (2005).

It should be noted, however, that only 7 HADS(B) were used for the model validation in the paper published by Petersen and Christensen-Dalsgaard, instead of 25 stars used in the paper by Poretti *et al.*

With regard to the relationship between the duration of the fundamental mode (P0) and the ratio between this and the first overtone (P1/P0), the first cited paper presents a model (their Figure 5) that predicts a peak value of the ratio around 0.774, which corresponds to log P0 = –0.9. For both increasing and decreasing duration of P0, the model predicts a lower ratio that becomes equal to 0.764 for values of log P0 = –0.55 and equal to 0.770 for log P0 = –1.1. In extreme simplification the model is similar to a downward parabola shape with the vertex (high ratio) for HADS with log P0 = –0.9.

In the paper published by Poretti *et al.* (their Figure 4) the model predicts a ratio characterized by a long standstill between log P0 = –1.30 and log P0 = –0.90. For the shortest period the ratio is increasing (0.778 for SX Phe itself) while for longer periods the ratio is decreasing (0.765 for GSC 04257-00471). In extreme synthesis the result of this model is to identify a direct relationship between the duration of the period and ratio (short periods have higher ratios while long periods have lower ratios) but considering the ratio stable for values between log P0 = –1.30 and log P0 = –0.90. The ratio's variability is mainly explained by lower metallicity for the higher values and in lower masses for the lower values.

On the other side I think it is important to mention the work of Pigulski *et al.* (2006) concerning the analysis of the data obtained from the OGLE-II (Udalski *et al.* 1997) and MACHO (Allsman and Axelrod 2001) projects. In this paper, the authors identify several other HADS(B) and publish a much more detailed Petersen diagram of the work here previously mentioned. However, while deciding to put in direct relation the duration of P0 and the ratio P1 / P0 (not a common choice as the diagram is normally realized with the logarithm of the fundamental period), they do not give any observations concerning a possible linear relation between the data nor, of course, its computed equation.

### 2. The Petersen diagram of HADS(B) stars

I decided to create a new Petersen's diagram using data from the AAVSO International Variable Star Index (Watson *et al.* 2014*)* related to 85 HADS(B), many of them completely unknown only five years ago. The period ratio was calculated by the author when not present in literature or simply reported when present. From these HADS(B) 8 stars were excluded for the following reasons:

• V798 Cyg: first and second overtone pulsator (Musazzi *et al.* 1998)

• V1719 Cyg: first and second overtone pulsator (Musazzi *et al.* 1998)

• VZ Cnc: first and second overtone pulsator (Fu and Jiang 1999)

• 1SWASP J211253.68+331734.3: probable second and third overtone pulsator (Khruslov 2014)

• ASAS J205850+0854.1: probable second and third overtone pulsator (Khruslov 2011)

• V1553 Sco: probable second and third overtone pulsator (Khruslov 2009)

• V526 Vel: probable second and third overtone pulsator (Khruslov 2011)

- V823 Cas: anomalous HADS because "the periods of the stars are in a transient, resonance affected state, thus do not reflect the true parameters of the object that is in effect a triple-mode pulsating variable." (Jurcsik *et al.* 2006)

The full list of stars used (as well as those not included) to create the Petersen's diagram is presented in Table 1, ordered by increasing P0.

The resulting diagram is shown in Figure 1, where the x-axis represents the log of P0 and the y-axis the ratio between P1 and P0.

Although the purpose of creating a new model capable of predicting the variation of the period and the ratio on the basis of the physical parameters of the star is outside the scope of this work, in observing the Petersen's diagram relative to all the HADS considered we may notice that the relationship between P0 and the period ratio does not appear as predicted by Petersen and Christensen-Dalsgaard (1996) and does not present even the long standstill described by Poretti *et al.* (2005).

Passing from an x-axis expressed as log P0 to an axis simply expressed in days, we notice that in fact the data seem well fitted by a straight line (red line), suggesting a possible linear relationship between the two factors as presented in Figure 2.

A greater scattering is certainly evident for the shortest periods and some stars are markedly outside the line of fit. However, considering the number of stars used for this plot

Table 1. List of stars used (as well as those not included) to create the Petersen's diagram, ordered by increasing P0.

| Name | R. A. (J2000) h m w | Dec. (J2000) ° ′ ″ | P0 duration (d) | P1 duration (d) | P1/P0 ratio |
|---|---|---|---|---|---|
| 2MASS J06451725+4122158 | 06 45 17.25 | +41 22 15.9 | 0.0500071 | 0.0386898 | 0.77369 |
| LINEAR 9328902 | 13 35 49.76 | +26 55 16.7 | 0.05174768 | 0.04046822 | 0.78203 |
| [SIG2010] 3269918 | 20 59 27.28 | –01 13 49.0 | 0.052376 | 0.040885 | 0.78061 |
| NSVS 10590484 | 15 13 22.01 | +18 15 58.3 | 0.0541911 | 0.0419105 | 0.77338 |
| USNO–B1.0 0961-0254829 | 15 52 51.38 | +06 06 06.1 | 0.05492 | 0.042667 | 0.77689 |
| SSS_J095657.2-231722 | 09 56 57.19 | –23 17 22.9 | 0.0566708 | 0.0442543 | 0.78090 |
| V879 Her | 17 31 12.72 | +28 03 16.8 | 0.0568926 | 0.044128 | 0.77564 |
| [MHF2014] J336.0969-15.6349 | 22 24 23.25 | –15 38 05.5 | 0.057182 | 0.044534 | 0.77881 |
| TSVSC1 TN-N231330220-6-67-2 | 08 58 54.72 | +15 22 09.7 | 0.0576289 | 0.044596 | 0.77385 |
| ASAS J061518+0604.2 | 06 15 17.73 | +06 04 12.6 | 0.0580806 | 0.044828 | 0.77182 |
| GSC 02008-00003 | 14 22 31.21 | +24 34 57.0 | 0.059596 | 0.046136 | 0.77415 |
| SDSS J151253.97+231748.4 | 15 12 53.99 | +23 17 48.3 | 0.06001381 | 0.0467412 | 0.77884 |
| GSC 07243-00871 | 12 08 49.77 | –36 33 11.1 | 0.060031 | 0.04648 | 0.77427 |
| BPS BS 16084-151 | 16 29 40.31 | +57 20 33.3 | 0.06114265 | 0.0475034 | 0.77693 |
| LINEAR 1683151 | 11 32 05.40 | –03 48 27.5 | 0.0618462 | 0.04820869 | 0.77949 |
| CSS_J213533.0+124341 | 21 35 32.99 | +12 43 41.3 | 0.0630537 | 0.0487775 | 0.77359 |
| NSVS 2577931 | 10 55 02.50 | +61 42 17.2 | 0.06404409 | 0.0496142 | 0.77469 |
| NSV 7805 | 16 32 20.12 | –02 12 08.3 | 0.064604 | 0.050699 | 0.78477 |
| OGLE BW2 V142 | 18 02 18.04 | –30 08 11.4 | 0.066041 | 0.051404 | 0.77836 |
| NSVS 2684702 | 13 45 21.66 | +54 11 51.2 | 0.06794351 | 0.0526002 | 0.77418 |
| SSS_J095011.1-244057 | 09 50 11.12 | –24 40 58.0 | 0.0683901 | 0.0530193 | 0.77525 |
| SEKBO 112944.737 | 20 10 22.51 | –23 10 59.7 | 0.0688009 | 0.0532926 | 0.77459 |
| LINEAR 16586778 | 16 13 57.55 | +28 28 57.2 | 0.070751 | 0.055701 | 0.78728 |
| V803 Aur | 06 12 13.90 | +31 48 24.4 | 0.0710556 | 0.0550312 | 0.77448 |
| FASTT 8 | 00 39 09.42 | +00 40 12.1 | 0.0730198 | 0.0571184 | 0.78223 |
| V1392 Tau | 04 26 05.90 | +01 26 26.2 | 0.07443025 | 0.05790307 | 0.77795 |
| KID 2857323 | 19 29 49.16 | +38 01 21.7 | 0.07618 | 0.05897 | 0.77409 |
| CSS_J214745.8+122726 | 21 47 45.78 | +12 27 26.6 | 0.07820144 | 0.06062011 | 0.77518 |
| [SIG2010] 2345453 | 21 29 52.69 | –01 10 18.9 | 0.080586 | 0.0624379 | 0.77480 |
| OGLE BW1 V207 | 18 02 14.98 | –29 54 08.8 | 0.085601 | 0.066234 | 0.77375 |
| MACHO 116.24384.481 | 18 13 16.45 | –29 49 27.0 | 0.086914 | 0.06716 | 0.77272 |
| GSC 07460-01520 | 20 33 38.54 | –32 55 03.6 | 0.087011 | 0.068152 | 0.78326 |
| NSVS 7293918 | 07 44 38.60 | +29 12 22.8 | 0.088535 | 0.068501 | 0.77372 |
| GSC 03693-01705 | 02 12 19.83 | +57 00 16.4 | 0.09108389 | 0.0704693 | 0.77367 |
| MACHO 115.22573.263 | 18 09 00.48 | –29 14 30.9 | 0.091754 | 0.070871 | 0.77240 |
| RV Ari | 02 15 07.46 | +18 04 28.0 | 0.0931281 | 0.0719466 | 0.77256 |
| QS Dra | 15 21 34.64 | +61 29 22.7 | 0.09442318 | 0.07304432 | 0.77358 |
| LINEAR 2653935 | 11 59 42.51 | +06 08 22.0 | 0.09520999 | 0.07460334 | 0.78357 |
| GSC 03949-00386 | 20 19 44.95 | +58 29 20.0 | 0.095783796 | 0.073937974 | 0.77193 |
| ASAS J094303-1707.3 | 09 43 02.81 | –17 07 15.9 | 0.0991782 | 0.07651564 | 0.77150 |
| USNO–A2.0 1425-12623576 | 21 59 23.24 | +59 24 56.9 | 0.1027306 | 0.079165 | 0.77061 |
| MACHO 114.19969.980 | 18 02 52.20 | –29 30 24.5 | 0.103272 | 0.079811 | 0.77282 |
| MACHO 119.19574.1169 | 18 02 00.37 | –29 48 43.2 | 0.1068464 | 0.082722 | 0.77421 |
| GSC 03887-00087 | 17 08 14.77 | +52 53 53.4 | 0.107183 | 0.082932 | 0.77374 |
| ASAS J182536-4213.6 | 18 25 36.26 | –42 13 35.8 | 0.1071934 | 0.0821611 | 0.76648 |
| [SIG2010] 2196466 | 21 36 30.17 | –00 21 27.6 | 0.107404 | 0.083675 | 0.77907 |
| BP Peg | 21 33 13.53 | +22 44 24.3 | 0.109543375 | 0.08451 | 0.77148 |
| V899 Car | 11 09 52.24 | –60 57 56.7 | 0.1108014 | 0.0858512 | 0.77482 |

Table 1. List of stars used (as well as those not included) to create the Petersen's diagram, ordered by increasing P0, cont.

| Name | R. A. (J2000) h m w | Dec. (J2000) ° ' " | P0 duration (d) | P1 duration (d) | P1/P0 ratio |
|---|---|---|---|---|---|
| MACHO 162.25343.874 | 18 15 16.33 | –26 35 40.2 | 0.111281 | 0.085905 | 0.77196 |
| AI Vel | 08 14 05.15 | –44 34 32.9 | 0.11157411 | 0.08620868 | 0.77266 |
| ASAS J231801-4520.0 | 23 18 01.14 | –45 19 55.0 | 0.1150105 | 0.0889176 | 0.77313 |
| 2MASS J18294745+3745005 | 18 29 47.55 | +37 45 01.5 | 0.116576 | 0.090297 | 0.77458 |
| V1393 Cen | 13 57 15.60 | –52 55 22.6 | 0.1177831 | 0.0908322 | 0.77118 |
| NSV 9856 | 17 56 00.20 | –30 42 46.6 | 0.118488 | 0.0912733 | 0.77032 |
| MACHO 128.21542.753 | 18 06 35.93 | –28 39 31.3 | 0.120052 | 0.09254 | 0.77083 |
| BPS BS 16553-0026 | 10 52 48.49 | +41 54 35.3 | 0.125508 | 0.096953 | 0.77248 |
| MACHO 114.19840.890 | 18 02 31.85 | –29 27 03.9 | 0.125566 | 0.096789 | 0.77082 |
| ASAS J152315-5603.7 | 15 23 15.43 | –56 03 43.2 | 0.1267467 | 0.0976718 | 0.77061 |
| GSC 04757-00461 | 05 23 54.48 | –03 07 32.3 | 0.1325305 | 0.1019376 | 0.76916 |
| GSC 02860-01552 | 03 16 02.70 | +43 20 34.3 | 0.13831414 | 0.10675322 | 0.77182 |
| V1384 Tau | 03 54 07.27 | +07 59 15.4 | 0.1397914 | 0.1073918 | 0.76823 |
| V575 Lyr | 18 29 43.24 | +28 09 54.6 | 0.1455591 | 0.1115016 | 0.76602 |
| ASAS J192227-5622.5 | 19 22 27.39 | –56 22 28.1 | 0.1490898 | 0.1127701 | 0.75639 |
| V703 Sco | 17 42 16.81 | –32 31 23.6 | 0.1499615 | 0.11521772 | 0.76832 |
| ASAS J062542+2206.4 | 06 25 41.61 | +22 06 19.5 | 0.1526484 | 0.117307 | 0.76848 |
| V403 Gem | 06 44 01.06 | +22 44 31.7 | 0.15338 | 0.117698 | 0.76736 |
| NSV 14800 | 00 01 16.22 | –60 36 57.1 | 0.1578385 | 0.122071 | 0.77339 |
| USNO–B1.0 1329-0132547 | 04 44 37.78 | +42 54 34.4 | 0.16189 | 0.12413 | 0.76676 |
| GSC 03949-00811 | 20 26 01.74 | +59 30 53.5 | 0.169751 | 0.1300791 | 0.76629 |
| GSC 04257-00471 | 21 26 01.11 | +64 30 57.5 | 0.173799 | 0.133084 | 0.76574 |
| V542 Cam | 04 53 46.52 | +68 28 26.5 | 0.174773 | 0.133986 | 0.76663 |
| DO CMi | 07 12 19.41 | +09 21 02.7 | 0.194506 | 0.14862 | 0.76409 |
| ASAS J194803+4146.9 | 19 48 02.92 | +41 46 55.8 | 0.203636 | 0.155488 | 0.76356 |
| VX Hya | 09 45 46.85 | –12 00 14.3 | 0.2233889 | 0.17272 | 0.77318 |
| V733 Pup | 08 18 06.98 | –22 14 07.7 | 0.2287147 | 0.1742342 | 0.76180 |
| AG Aqr | 22 05 31.82 | –22 30 00.7 | 0.291736 | 0.2222 | 0.76165 |
| V829 Aql | 19 46 57.29 | +03 30 28.5 | 0.292444 | 0.220972 | 0.75560 |
| *Stars excluded* | | | | | |
| V798 Cyg | 19 38 06.90 | +30 54 33.5 | | | |
| V1719 Cyg | 21 04 32.92 | +50 47 03.3 | | | |
| VZ Cnc | 08 40 52.12 | +09 49 27.2 | | | |
| V823 Cas | 00 05 42.38 | +63 24 14.2 | | | |
| 1SWASP J211253.68+331734.3 | 21 12 53.69 | +33 17 34.3 | | | |
| ASAS J205850+0854.1 | 20 58 49.64 | +08 54 05.3 | | | |
| V1553 Sco | 16 20 21.77 | –35 41 16.0 | | | |
| V526 Vel | 09 03 13.34 | –52 02 28.7 | | | |

(almost 3 times compared to the works of Poretti *et al.*) we can actually note that the greatest number of stars lie along the path of the fit line. This evidence is also highlighted by looking at the ratio residuals compared to the best-fit linear regression (the computed equation is $Y = -0.084809X + 0.782048$) as presented in Figure 3.

In this case even a second and third polynomial fit of the residuals (red and blue lines) shows a substantial absence of trend, suggesting that a linear interpretation of the relation is possible, a fact that argues for a more accurate revision of HADS stellar models than so far proposed in the literature.

From a purely theoretical point of view I suggest this interpretation: in short period stars the metallicity could vary greatly simply because double mode pulsators with such short periods are, for example, characteristic of double-mode SX Phe stars, characterized precisely by low metallicity as explained in McNamara (2000). In other words the area with period shorter than 0.1 day is probably a transition area with stars of population I and II mixed together, and thus the stellar parameters are less homogeneous than in typical double-mode HADS. This could result in a stronger scattering that does not, however, affect the linear relation suggested.

Finally, the two stars with the highest residuals in the diagram relative to period > 0.1 day (ASAS J192227-5622.5 and VX Hya) could be stars for which the pure nature of double-mode HADS should be evaluated more carefully, as was the case for the previously cited V798 Cyg. For example, VX Hya was involved in a careful analysis by an AAVSO campaign in 2006 and 2007 and the data obtained (Templeton *et al.* 2009) showed that it is certainly an HADS(B) but characterized by unusual and not fully explained long-term amplitude variations.

The linear relationship proposed in this paper could then more easily show which stars belong to a pure type HADS(B): the presence of unusual peculiarity immediately puts the star clearly outside the best-fit line.

Of course the presence of a greater number of stars identifiable as Double-Mode HADS could significantly improve the results of this work in determining the correct parameters of this possible linear relation. I believe that much work can be done from the large amount of data collected from large photometric surveys of recent years (OGLE, SuperWASP, ASAS, and so on) and from the data of exceptional quality obtained from the Kepler satellite. This can certainly be a stimulus for new and extensive works.

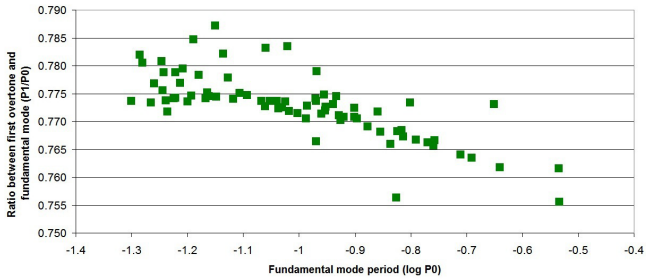

Figure 1. Double Mode HADS Petersen diagram with fundamental mode period expressed as log P0.

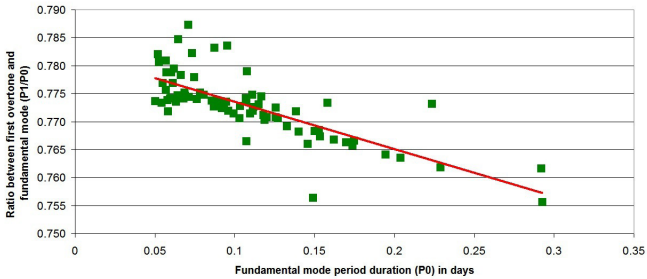

Figure 2. Double Mode HADS Petersen diagram with fundamental mode period expressed in days. The red line represent the best-fit linear regression.

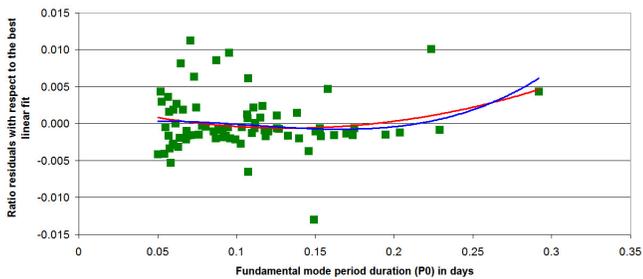

Figure 3. P1/P0 ratio residuals for the best-fit linear regression vs period duration. The red and blue lines represent, respectively, a second and third degree polynomial fit showing no relevant residual trends of the data.

### 3. Fit line parameters and statistical correlation evidence

The calculated equation for the best-fit line presented in Figure 2 is:

$$Y = -0.084809 \, (\pm 0.008298) X + 0.782048 \, (\pm 0.000995) \quad (1)$$

with an RMS error = 0.003765 and a correlation coefficient = 0.762926.

### 4. Acknowledgements

This work has made use of the International Variable Star Index (VSX; Watson *et al.* 2014) operated by the AAVSO, Cambridge, Massachusetts, USA. The best fit line and relative errors were calculated with the Nonlinear Least Squares Regression (Curve Fitter) courteously provided by Prof. John C. Pezzullo at http://statpages.org/nonlin.html.